\documentclass[twocolumn,showpacs,amsmath,amssymb]{revtex4}

\usepackage{graphicx}
\usepackage{dcolumn}
\usepackage{bm}

\begin{document}


\title{Riemannian light cone from vanishing birefringence \\ 
in premetric vacuum electrodynamics}

\author{Claus L{\"a}mmerzahl} \email{laemmerzahl@zarm.uni-bremen.de}
\affiliation{Centre of Advanced Space Technology and Microgravity (ZARM), 
University of Bremen, Am Fallturm, 28359 Bremen, Germany}%

\author{Friedrich W.\ Hehl} \email{hehl@thp.uni-koeln.de}
\affiliation{ Institute for Theoretical Physics, University of
  Cologne, 50923 K\"oln, Germany \\ and \\ 
Department of Physics and Astronomy, University of Missouri-Columbia,\\ Columbia, MO 65211, USA}%

\date{15 November 2004, {\it file lightcone14a.tex}}

\begin{abstract}
  We consider premetric electrodynamics with a local and linear
  constitutinen7vareve law for the vacuum. Within this framework, we find
  quartic Fresnel wave surfaces for the propagation of light. If we
  require (i) the Fresnel equation to have only real solutions and
  (ii) the vanishing of birefringence in vacuum, then a Riemannian
  light cone is implied. No proper Finslerian structure can occur.
  This is generalized to dynamical equations of any order.
\end{abstract}

\maketitle

\section{Introduction and Motivation}

Recently, the physics of the electromagnetic field, without
assumptions about the metrical structure of the underlying spacetime,
has gained renewed interest. On the one hand, this general ansatz is
needed for a proper interpretation of experiments testing Lorentz
invariance. In such approaches it is not allowed to make assumptions
about the underlying geometric structure, in particular, about a
metric of spacetime. On the contrary, by using properties of the
evolution of the electromagnetic field, one likes to establish the
metrical structure of spacetime (here ``metrical'' may be more general
than the ordinary Riemannian or Minkowskian metric). The general
structure of Maxwell equations can serve as a test theory for searches
for Lorentz violation in the photon sector \cite{KosteleckyMewes02}.
On the other hand, it is a general task to explore the structure of
the electromagnetic field and the geometry it defines, see e.g.\ 
\cite{HehlObukhov03,LaemmerzahlMaciasMueller04}.

There are two main effects in the realm of ray optics based on the
Maxwell equations: One effect is {\it birefringence\/} and the other
one {\it anisotropy\/} of the propagation of light \footnote{A
further propagation effect of electromagnetic radiation is the
precession of its polarization. This will not be discussed in this
paper.}. Both effects are well known from the physics of light
propagation in general media, such as in crystals, e.g.. The basics
of the general formalism have been laid down in
\cite{HehlObukhov03}. The explicit calculations of these effects
have been carried through to first order in these anomalous effects
by Kostelecky and coworkers and by others
\cite{Ni74,Ni84,HauganKauffmann95,KosteleckyMewes02,LaemmerzahlMaciasMueller04}
(for a possible birefringence caused by a torsion of spacetime, see
\cite{Solanki,Preuss:2004pp,Rubilar:2003uf,Itin:2003hr} and also
\cite{Itin:2004za}). In these approaches the first step is to
confront the result with the possible observations of birefringence.
{}From astrophysical observations \cite{KosteleckyMewes02}, the
parameters responsible for birefringence must be smaller than
$10^{-32}$ and, thus, can safely be neglected. The remaining
anisotropy in the photon propagation is given by a symmetric
second--rank tensor. By an appropriate coordinate transformation,
this tensor becomes proportional to the unit tensor. Accordingly,
there is an adapted coordinate system such that light propagation is
isotropic and defines a Riemannian metric. This is a remarkable
result that may be due to the approximation used. In this work we
show that this result holds exactly. That is, we show, provided we
assume a local and linear constitutive law for the vacuum, that
\begin{equation*}
  \left. \begin{array}{l} \hbox{\;\;\;Maxwell equations} \\ \hbox{+\;only
        real sols.\ of Fresnel eq.} \\ \hbox{+\;vanish.\
        birefringence in vac.} \end{array} \right\} \;\, \Rightarrow\;\,
  \begin{matrix} \hbox{pseudo-Riemann-} \\ \hbox{ian metric \, .}
  \end{matrix}
\end{equation*}

\section{Observational and experimental facts}

As discussed, the best estimate on birefringence effects of the vacuum
have been given by an analysis of Kostelecky and Mewes
\cite{KosteleckyMewes02}. Their results show that the birefringence
parameter is smaller than $10^{-32}$. This estimate is independent of
the coordinate system chosen since it is an effect which cannot be
transformed to zero.

Since the resulting anisotropy can be transformed away, it cannot be
understood as an effect solely within the photon sector. Thus, the
coordinates used for the description of the anisotropy experiments
have to be fixed by some other physical process. In these experiments,
this is realized by some solid like, e.g., the interferometer arm or
the optical resonator. The length of the resonator or of the
interferometer arm is determined by the laws of quantum physics and of
electrodynamics, see \cite{Muelleretal03,Muelleretal03d}, e.g..
Consequently, the search for an anisotropy of the propagation of light
has to be interpreted as a comparison between the laws of quantum
physics, like the Dirac equation, the Pauli exclusion principle, etc.,
and the Maxwell equations. The most recent experiments searching for
an anisotropy of the velocity of light yield no effect to the order of
$\Delta c/c \leq 10^{-15}$ \cite{Muelleretal03c}.

\section{Some premetric electrodynamics}

The Maxwell equations, expressed in terms of the excitations ${\cal
  D},{\cal H}$ and the field strengths $E,B$, read
\begin{eqnarray}\label{Maxwell}
  \underline{d}\,{\cal D}&=&\rho\,,\qquad \underline{d}\,{\cal
    H}-\dot{{\cal D}}=j\,, \\ \underline{d}\,{
    B}&=&0\,,\qquad\underline{d}\,{E}+\dot{{B}}=0\,.
\end{eqnarray}
We mark the exterior derivative in 3 dimensions with an underline:
$\underline{d}$. The dot denotes a Lie derivative with respect to
the vector field $\partial_t\,$. The electric charge density is
$\rho$, the current density $j$. For the formulation of the Maxwell
equations, we use the calculus of exterior differential forms. We
take the notation from \cite{HehlObukhov03}, compare also Frankel
\cite{Frankel}, Lindell \cite{Lindell}, or Russer \cite{Russer},
e.g..

The 4 dimensional form of the Maxwell equations
\begin{eqnarray}
  d\,H&=&J\,,\quad H = {\cal D}-\,{\cal H}\wedge dt \,,\quad J =\rho
  -\,j\wedge dt\, \,,\label{inhMax}\\ d\,F&=&\,0\,,\quad\, F
  =B+\,E\wedge dt \,,\label{homMax}
\end{eqnarray}
shows that they are generally covariant under diffeomorphisms and
there is no need of a metric of spacetime \cite{HehlObukhov03}.

The set of equations (\ref{inhMax}) and (\ref{homMax}) is incomplete.
What is missing is the constitutive law of the vacuum (the spacetime
relation). If we assume {\it locality} and {\it linearity,} then
$H=\kappa(F)$, with the local and linear operator $\kappa$. If we
decompose the 2--forms $H$ and $F$ in their components (here
$i,j=0,1,2,3$), then $H=H_{ij}\,dx^i\wedge dx^j/2$ and
$F=F_{ij}\,dx^i\wedge dx^j/2$. Accordingly,
\begin{equation}\label{constit} H_{ij} =
  \frac{1}{2}\,\kappa_{ij}{}^{kl}\,F_{kl}\quad {\rm with}\quad
  \kappa_{ij}{}^{kl}=-\, \kappa_{ji}{}^{kl} =-\,\kappa_{ij}{}^{lk}.
\end{equation}
Here $\kappa_{ij}{}^{kl}$ is the constitutive tensor of spacetime with
36 independent components. With the help of the contravariant
Levi--Civita symbol ${\epsilon}^{ijmn}=\pm 1,0$, we can introduce the
equivalent constitutive tensor {\em density} of spacetime,
\begin{equation}\label{raise}
  \chi^{ijkl}:= \frac{1}{2}\,{\epsilon}^{ijmn}\, \kappa_{mn}{}^{kl}
  \,.
\end{equation}
Incidentally, the covariant Levi-Civita symbol, which we will use
below, is denoted by a circumflex: $\hat{{\epsilon}}_{ijmn}=\pm 1,0$.
Since no metric is available at this stage, we have to distinguish
these two symbols.

Alternatively, we can express (\ref{constit}) in a six component
version, which is sometimes more convenient. In terms of blocks with
3--dimensional indices $a,b,\dots =1,2,3$, we find
\begin{equation}
  \left(\begin{array}{c} {\cal H}_a \\ {\cal D}^a\end{array}\right) =
  \left(\begin{array}{cc} {{\frak C}}^{b}{}_a & {{\frak B}}_{ba} \\
      {{\frak A}}^{ba}& {{\frak D}}_{b}{}^a \end{array}\right)
  \left(\begin{array}{c} -E_b\\ {B}^b\end{array}\right)\,.\label{CR'}
\end{equation}
Obviously, ${\frak A}$ is the 3--dimensional {\it permittivity} matrix
and ${\frak B}$ the reciprocal of the {\it permeability} matrix. The
matrices ${\frak C}$ and ${\frak D}$ describe electric-magnetic cross
terms (which vanish in Maxwell--Lorentz vacuum electrodynamics in
Cartesian coordinates). In (\ref{CR'}), for the components of the
electromagnetic field, we took a vector-like notation
\begin{eqnarray}
  {\cal H}&=& {\cal H}_a\, \vartheta^a\,, \quad E=E_a\, \vartheta^a\,,
  \\ {\cal D}&=& { {\cal D}}^b\,\, \hat{\epsilon}_b\,,\quad\>
  B={B}^b\, \hat{\epsilon}_b\,,
\end{eqnarray}
with the 3--dimensional coframe $\vartheta^a$ and the 2--form basis
$\hat{\epsilon}_a={\hat{\epsilon}}_{bcd}\,\vartheta^c\wedge
\vartheta^d/2$. By straightforward algebra, the constitutive $3\times
3$ matrices $\frak{A},\frak{B},\frak{C},\frak{D}$ can be related to
the 4-dimensional constitutive tensor density (\ref{raise}) by
\begin{eqnarray}\label{A-matrix0}
  {\frak A}^{ba}&:=& \chi^{0a0b}\,,\\
\label{B-matrix0} {\frak B}_{ba}&:=&\frac{1}{4}\,\hat\epsilon_{acd}\,
\hat\epsilon_{be\!f} \,\chi^{cdef}\,,\\
\label{C-matrix0}
{\frak C}^a{}_b& :=&\frac{1}{2}\,\hat\epsilon_{bcd}\,\chi^{cd0a}\,,\\
\label{D-matrix0} {\frak D}_a{}^b&:=&\frac{1}{2}\,\hat\epsilon_{acd}
\,\chi^{0bcd}\,.
\end{eqnarray}

\section{Quartic wave surface for the propagation of light}

The propagation of light in {\it local} and {\it linear\/} premetric
vacuum electrodynamics is characterized by the generalized Fresnel
equation \cite{HehlObukhov03}
\begin{equation}\label{Fresnel}
  M_0 k_0^4 + M_1 k_0^3 + M_2 k_0^2 + M_3 k_0 + M_4 = 0\,,
\end{equation}
where $k_0$ is the zeroth component of the 4--wave covector $k$. The
coefficients $M_i$ are homogeneous functions of degree $i$ in the
spatial components $k_a$ of the wave covector:
\begin{equation}\label{emms}
  M_i := M^{a_1\ldots a_i} k_{a_1}\cdots k_{a_i}\,.
\end{equation}
The Fresnel equation results from a solvability condition for a
3--vector equation $W^{ab} k_b=0$ on the jump surfaces
\cite{HehlObukhov03,Obukhov:2000nw}; here 
\begin{eqnarray}
W^{ab} & := & \left(k_0^2\,{\frak
    A}^{ab} + k_0k_d\left[{\frak C}^a{}_c\,\epsilon^{cdb} +{\frak
      C}^b{}_c\,\epsilon^{cda}\right] \right. \nonumber\\
& & \left. \qquad  +k_ek_f\,
  \epsilon^{aec}\epsilon^{bfd}\,{\frak B}_{cd}\right)
\end{eqnarray}
is a $3\times 3$
matrix, the determinant of which has to vanish, see
\cite{Obukhov:2000nw}. Eq.\ (\ref{emms}) is valid in a special
anholonomic frame with $\vartheta^0=k$.

The equation for the jump surfaces can also be obtained in an
analogous way as effective partial differential equation for the
components of the radiating electromagnetic potential after removing
all gauge freedom. This equation, for all initial conditions or all
sources of sufficient regularity, should possess a unique solution
in some future causality cone (this corresponds to a finite
propagation velocity of the solutions). The necessary and sufficient
condition for that is the hyperbolicity of the differential operator
\cite{Hoermander}. Furthermore, the differential operator is
hyperbolic if the corresponding polynomial is hyperbolic
\cite{Hoermander}. This means that (\ref{Fresnel}) is required to
possess four {\it real\/} solutions for $k_0$ which need not to be
different. The condition of the hyperbolicity or, equivalently, the
condition for the existence and the uniqueness of the solutions, is
the fundamental fact behind the particular signature for the metric
which we are going to derive (see also \cite{AudretschLaemmerzahl93}
for another example).

The $M^{a_1\dots a_i}$'s in (\ref{emms}) are cubic in the $3\times
3$ matrices $\frak A$, $\frak B$, $\frak C$, and $\frak D$, see
\cite{HehlObukhov03}:
\begin{widetext}
\begin{eqnarray}  \label{ma0}
  M&=&\det{\frak A} \,, \\ M^a&=& -\hat{\epsilon}_{bcd}\left( {\frak
      A}^{ba}\,{\frak A}^{ce}\, {\frak C}^d_{\ e} + {\frak
      A}^{ab}\,{\frak A}^{ec}\,{\frak D}_e^{\ d}
  \right)\,,\label{ma1}\\ M^{ab}&=& \frac{1}{2}\,{\frak
    A}^{(ab)}\left[({\frak C}^d{}_d)^2 + ({\frak D}_c{}^c)^2 - ({\frak
      C}^c{}_d + {\frak D}_d{}^c)({\frak C}^d{}_c + {\frak
      D}_c{}^d)\right] +({\frak C}^d{}_c + {\frak D}_c{}^d)({\frak
    A}^{c(a}{\frak C}^{b)}{}_d + {\frak D}_d{}^{(a}{\frak A}^{b)c})
  \nonumber\\ && - {\frak C}^d{}_d {\frak A}^{c(a}{\frak C}^{b)}{}_c -
  {\frak D}_c{}^{(a}{\frak A}^{b)c}{\frak D}_d{}^d - {\frak
    A}^{dc}{\frak C}^{(a}{}_c {\frak D}_d{}^{b)} + \left({\frak
      A}^{(ab)}{\frak A}^{dc}- {\frak A}^{d(a}{\frak
      A}^{b)c}\right){\frak B}_{dc}\,,\label{ma2}\\ M^{abc} &=&
  \epsilon^{de(c|}\left[{\frak B}_{df}( {\frak A}^{ab)}\,{\frak
      D}_e^{\ f} - {\frak D}_e^{\ a}{\frak A}^{b)f}\,) + {\frak
      B}_{fd}({\frak A}^{ab)}\,{\frak C}_{\ e}^f - {\frak
      A}^{f|a}{\frak C}_{\ e}^{b)}) +{\frak C}^{a}_{\ f}\,{\frak
      D}_e^{\ b)}\,{\frak D}_d^{\ f} + {\frak D}_f^{\ a}\,{\frak
      C}^{b)}_{\ e}\,{\frak C}^{f}_{\ d} \right] \, ,\label{ma3}\\ 
  M^{abcd} &=& \epsilon^{e\!f(c}\epsilon^{|gh|d}\,{\frak B}_{hf}
  \left[\frac{1}{2} \,{\frak A}^{ab)}\,{\frak B}_{ge} - {\frak
      C}^{a}_{\ e}\,{\frak D}_g^{\ b)}\right] \,.\label{ma4}
\end{eqnarray}
\end{widetext}
Computer plots of the 4th-order surface of the generalized Fresnel
equation (\ref{Fresnel}) have been prepared by Tertychniy
\cite{Sergey}.

We solve (\ref{Fresnel}) with respect
to the frequency $k_0$, keeping the 3--covectors $k_a$ fixed. We find
the four solutions
\begin{eqnarray}\label{freqk1+}
k_{0(1)}^{\hspace{7pt}\uparrow} & = & \hspace{8pt}\sqrt{\alpha} + \sqrt{\beta
+ \frac{\gamma}{\sqrt{\alpha}}}
 - \delta\, , \\ \label{freqk2+}
k_{0(2)}^{\hspace{7pt}\uparrow} & = & \hspace{8pt}\sqrt{\alpha} - \sqrt{\beta
+ \frac{\gamma}{\sqrt{\alpha}}}
 - \delta\, ,\\ \label{freqk1-}
k_{0(1)}^{\hspace{7pt}\downarrow} & = & -\sqrt{\alpha} + \sqrt{\beta
-\frac{\gamma}{\sqrt{\alpha}}}
 - \delta\, ,\\ \label{freqk2-}
k_{0(2)}^{\hspace{7pt}\downarrow} & = & -\sqrt{\alpha} - \sqrt{\beta
-\frac{\gamma}{\sqrt{\alpha}}}
 - \delta \,.
\end{eqnarray}
We introduced the abbreviations
\begin{eqnarray}
\alpha & := &  \frac{1}{12M_0}\left(\hspace{8pt}\frac{ a}{\left(b +
\sqrt{c}\right)^{\frac{1}{3}}}
 + \left(b + \sqrt{c}\right)^{\frac{1}{3}} - 2M_2\right)+\delta^2\,, \nonumber\\ 
\\
\beta & := &  \frac{1}{12M_0}\left(-\frac{ a}{\left(b +
\sqrt{c}\right)^{\frac{1}{3}}}
 - \left(b + \sqrt{c}\right)^{\frac{1}{3}} - 4M_2\right)+2\delta^2\,, \nonumber\\
\\
\gamma & := & \frac{1}{4M_0}\left(2\delta M_2-M_3 \right)-2\delta^3\,,\\
\delta&:=&\frac{M_1}{4M_0}\,,
\end{eqnarray}
with
\begin{eqnarray}
a & := &    12 M_0 M_4 - 3 M_1 M_3 +M_2^2 \,,\\
b & := & \frac{27}{2} M_0 M_3^2- 36 M_0 M_2 M_4 \nonumber\\
& & - \frac{9}{2} M_1 M_2
M_3+
\frac{27}{2} M_1^2 M_4 +M_2^3 \,, \\
c & := & {4}\left(b^2 - a^3\right)\, .
\end{eqnarray}

Earlier investigations on light propagation in general linear media and on
Fresnel-Kummer surfaces includes the important work of Schultz et al.\ 
\cite{Schultz} and Kiehn et al.\ \cite{KiehnFresnel}.

\section{Vanishing birefringence}

Vanishing birefringence means that there is only one future and only
one past directing light cone. In order to achieve this, one has to
identify two pairs of solutions. There are these two
possibilities \footnote{For $\gamma=\delta=0$, we have
  $k_{0(1)}^{\hspace{7pt}\uparrow} =-
  k_{0(2)}^{\hspace{7pt}\downarrow}$ and
  $k_{0(2)}^{\hspace{7pt}\uparrow} =-
  k_{0(1)}^{\hspace{7pt}\downarrow}$. However, this is irrelevant for
  birefringence.}:
\begin{eqnarray}\label{case1}
k_{0(1)}^{\hspace{7pt}\uparrow} =
  k_{0(2)}^{\hspace{7pt}\uparrow}\,,\quad
  k_{0(1)}^{\hspace{7pt}\downarrow} =
  k_{0(2)}^{\hspace{7pt}\downarrow}\,,&&\;\text{i.e.,}\quad \beta=\gamma=0\,, \qquad\\
\label{case2} k_{0(1)}^{\hspace{7pt}\uparrow} =
  k_{0(1)}^{\hspace{7pt}\downarrow} \,,\quad
  k_{0(2)}^{\hspace{7pt}\uparrow} =
  k_{0(2)}^{\hspace{7pt}\downarrow}\,,&&\;\text{i.e.,}\quad \alpha=\gamma=0\,.
\end{eqnarray}
For the case (\ref{case1}), the solution degenerates to
\begin{equation}
k_0^\uparrow = \sqrt{\alpha}  - \delta \, , \qquad k_0^\downarrow = - \sqrt{\alpha}
- \delta \, ,
\end{equation}
and for the case (\ref{case2}) to
\begin{equation}
k_0^\uparrow = \sqrt{\beta}  - \delta \, , \qquad
k_0^\downarrow = - \sqrt{\beta}  - \delta \, .
\end{equation}

The equation $\gamma = 0$, which is valid for both cases, has the
simple solution
\begin{equation}
M_3 = \frac{M_1 M_2}{2 M_0} - \frac{1}{8} \frac{M_1^3}{M_0^2}
=\frac{M_1}{8M_0^2}\left(4M_0M_2-M_1^2 \right)\, .
\end{equation}
This can be inserted into $a$ and $b$, but presently we don't need the
explicit expressions. The functions $\alpha$ and $\beta$ can be written as
\begin{eqnarray}
  \alpha & = & \frac{3 M_1^2 - \;8 M_0 M_2}{48 M_0^2} \,+ \xi
  \label{ConditionA}\,, \\ \beta & = & \frac{6 M_1^2 - 16 M_0 M_2}{48
    M_0^2} - \xi\,,
\label{ConditionB}\vspace{-10pt}
\end{eqnarray}
with
\begin{equation}\label{AforB=0}
\xi:=  \frac{1}{12M_0}\left(\hspace{8pt}\frac{ a}{\left(b +
\sqrt{c}\right)^{\frac{1}{3}}}
 + \left(b + \sqrt{c}\right)^{\frac{1}{3}}\right)\,.
 \end{equation}
Since either $\beta = 0$ or $\alpha = 0$, we can add (\ref{ConditionA}) and
 (\ref{ConditionB}) and find
\begin{equation}
\alpha \quad\hbox{or} \quad \beta = \frac{3M_1^2 - 8 M_0 M_2}{16 M_0^2}\,,
\end{equation}
corresponding to (\ref{case1}) or to (\ref{case2}), respectively.

Hence in all cases the light cones turn out to be
\begin{equation}\label{solut1}
  k_0^{\uparrow\!\downarrow}=\pm\sqrt{\frac{3M_1^2-8M_0M_2}{16M_0^2}}-
  \frac{M_1}{4M_0}\,.
\end{equation}
Accordingly, the quartic wave surface in this case reads
\begin{equation}\label{solut2}
[(k_0-k_0^{\uparrow})(k_0-k_0^{\downarrow})]^2=0\,.
\end{equation}
We drop the square and find
\begin{eqnarray}\label{solut3}
\left(k_0+ \frac{M_1}{4M_0}
-\sqrt{\frac{3M_1^2-8M_0M_2}{16M_0^2}}\right) \qquad\qquad\qquad & & \nonumber \\ 
\times \left(k_0+\frac{M_1}{4M_0}+
\sqrt{\frac{3M_1^2-8M_0M_2}{16M_0^2}}\right)=0 \,. & & 
\end{eqnarray}
Multiplication yields
\begin{equation}\label{solut4}
\left( k_0+
\frac{M_1}{4M_0}\right)^2-\frac{3M_1^2-8M_0M_2}{16M_0^2}=0\,
\end{equation}
or
\begin{equation}\label{solut5}
 k_0^2+\frac 12\,\frac{M_1}{M_0}\,k_0+\frac 12\,\frac{M_2}{M_0}
 -\frac 18\,\left(\frac{M_1}{M_0}\right)^2 =0 \,.
\end{equation}
If we substitute the $M_i$'s according to (\ref{emms}), we have
\begin{eqnarray}\label{solut6}
g^{ij} k_i k_j & := & k_0^2+\frac 12\,\frac{M^a}{M}\,k_0k_a \nonumber\\
& & +\frac
18\left(4\,\frac{M^{ab}}{M}
 -\frac{M^aM^b}{M^2}\right)k_ak_b \nonumber\\
& = & 0 \,.
\end{eqnarray}
This form is quadratic in the wave 4--covector $k_i$ and thus
constitutes, up to a scalar factor, a Riemannian metric. Equation
(\ref{solut6}) represents our main result%
. It is clear that there is a coordinate system so that the metric
$g^{ij}$ acquires the ordinary Minkowski form: $g^{ij}
\overset{*}{=} {\rm diag}(+1,-1,-1,-1)$. Therefore, intrinsically it
is not possible to have an anisotropic speed of light.

{}From the condition of the existence of a unique solution (or from
hyperbolicity), equation (\ref{solut6}) has to possess two real
solutions for any given spatial $k_a$. As a consequence, the
signature of the metric $g^{ij}$ is $(+1,-1,-1,-1)$. Accordingly,
the signature of the metrical structure is a consequence of the
existence of a unique solution of the Maxwell equations in a future
causal cone for arbitrary sources with compact support.

Let us look at a specific example. If we exclude, besides
birefringence, also {\it electric-magnetic cross terms\/} in the
spacetime relation (\ref{CR'}), then ${\frak C}={\frak D}=0$ and,
according to (\ref{ma1}), $M^a=0$. If we substitute this into
 (\ref{solut6}), we find
\begin{equation}\label{solut7}
{k_0}^2+\frac{M^{ab}k_ak_b}{2M}=0\,.
\end{equation}
It can be shown \cite{forerunner} that one arrives also at this result
by only forbidding the existence of electric-magnetic cross terms,
that is, this condition is stronger than the requirement of vanishing
birefringence. Clearly then, for the Minkowskian signature we have
\begin{equation}
\frac{M^{ab}k_ak_b}{2M}<0\,,
\end{equation}
see also (\ref{solut6}). The flat Minkowski spacetime of special
relativity is a subcase of (\ref{solut7}). Then, in Cartesian
coordinates, $M^{ab}$ is a constant. This is a consequence of the
constancy of the constitutive matrices ${\frak A}^{ba}$ and ${\frak
B}_{ba}$. Because of (\ref{CR'}), we find ${\cal D}^a=-{\frak
A}^{ba}E_b$ and ${\cal H}_a={\frak B}_{ba}B^b$. Thus,
\begin{equation}
{\frak A}=-\varepsilon_0\,\mathbf{1}_3\,,\quad {\frak
B}=\frac{1}{\mu_0}\,\mathbf{1}_3\,,
\end{equation}
where $\mathbf{1}_3$ denotes the 3--dimensional unit matrix. If we
substitute this into (\ref{ma0}) to (\ref{ma2}), we find
$M=-\varepsilon_0^3$, $\,M^a=0$, and
$\,M^{ab}=(2{\varepsilon_0}^2/\mu_0)\,\mathbf{1}^{ab}$, that is,
\begin{equation}
\frac{M^{ab}}{2M}=-\frac{1}{\varepsilon_0\mu_0}\,\mathbf{1}^{ab}=
-{c^2}\,\mathbf{1}^{ab}
\end{equation}
is negative, with $c$ as the speed of light in vacuum.

Note that the vanishing of birefringence is not equivalent to the
validity of the reciprocity relation as discussed in
\cite{HehlObukhov03}.

\section{A unique light cone is incompatible with a
Finslerian geometry}\label{Finsler}

Now we would like to generalize the result obtained above: {\em For
all hyperbolic partial differential equations a vanishing
birefringence of the characteristic cones defines merely a
Riemannian structure}. There is no way to have two characteristic
cones with a Finslerian structure. In fact, the restriction to
hyperbolic partial differential equations is necessary for physical
reasons: only for hyperbolic partial differential equations one has
a unique solution in the future half space for prescribed initial
values or prescribed source, see \cite{Hoermander}.

Let us now prove the above statement: Any characteristic surface is
given by a polynomial of order $p$ in the covector $k_i$, which is
`normal' to the characteristic surface \footnote{Strictly, a covector
  or 1--form is visualized by two parallel planes. If $\phi=0$
  describes the jump surface, then $k_i=\partial_i
  \phi$. Thus the two planes visualizing the 1--form are parallel to
  the tangent plane of the jump surface.},
\begin{equation}
H(k) = \mathfrak{g}^{i_1 i_2 \ldots i_p} k_{i_1} k_{i_2} \cdots
k_{i_p}  \,.
\end{equation}
In order to be based on a hyperbolic differential operator, this
polynomial also has to be hyperbolic, that is, there should exist $p$
real solutions $k_0 = k_0(k_a)$ (see, e.g., \cite{Hoermander})
\begin{equation}
H(k) = \prod_{m=0}^p \left(k_0 - k_{0(m)}\right) \, .
\end{equation}
This specifies a splitting of the characteristic cone into $p$
sheets.

Now we want to restrict the number of cones to 2. In order to be
able to identify an equal number of cones, we choose $p = 2 q$.
After the identification of the first $q$ solutions and the last $q$
solutions, respectively, we have as characteristic polynomial
\begin{eqnarray}
H(k) & = & \mathfrak{g}^{i_1 i_2 \ldots i_{2q}} k_{i_1} k_{i_2} \cdots
k_{i_{2q}} \nonumber\\ 
& = & \left(k_0 - k_{0(1)}\right)^q \left(k_0 -
k_{0(2)}\right)^q \nonumber\\
& = & \left[k_0^2 - \left(k_{0(1)} + k_{0(2)}\right) k_0 + k_{0(1)}
k_{0(2)}\right]^q  , \label{Riemlightcone}
\end{eqnarray}
where the two solutions $k_{0(1)}$ and $k_{0(2)}$ are homogeneous
functions of the spatial components $k_a$.

We differentiate this relation with respect to $k_a$ and set
subsequently $k_a = 0$. This results in $k_{0(1,2)}(k_a = 0) = 0$.
For the zeroth derivative we get
\begin{equation}
\mathfrak{g}^{0 0 \ldots 0} = 1\,.
\end{equation}
The first derivative reads
\begin{eqnarray}
&& 2 n \mathfrak{g}^{i_1 \cdots i_{2q - 1} a} k_{i_1} \cdots k_{i_{2q -
1}} = \nonumber\\
& & \qquad q \left(k_0^2 - \left(k_{0(1)} + k_{0(2)}\right) k_0 +
k_{0(1)}
k_{0(2)}\right)^{q-1} \times \nonumber\\
& & \qquad \left(- \frac{\partial}{\partial k_a}\left(k_{0(1)} +
k_{0(2)}\right) k_0 + \frac{\partial}{\partial k_a}\left(k_{0(1)}
k_{0(2)}\right)\right) , \qquad 
\end{eqnarray}
which, for $k_a \rightarrow 0$, yields
\begin{eqnarray}
2 \mathfrak{g}^{(0 \ldots 0 a)} & = & - \frac{\partial}{\partial
k_a}\left(k_{0(1)} + k_{0(2)}\right)\,.
\end{eqnarray}
This can be integrated to
\begin{equation}
  k_{0(1)} + k_{0(2)} = - 2 \mathfrak{g}^{(0 \ldots 0 a)} k_a =: -2
  g^{0a} k_a
\label{solk1+k2}
\end{equation}
(no constant must be added because the $k_{0(m)}$'s are homogenous in
$k_a$).

Analogously, we calculate the second derivative and perform the limit
$k_a \rightarrow 0$,
\begin{eqnarray}
  2 (2 q - 1) \mathfrak{g}^{(0 \ldots 0 ab)} & = & 4 \mathfrak{g}^{(0
    \ldots 0 a)} \mathfrak{g}^{(0 \ldots 0 b)} \nonumber\\
& & +
  \frac{\partial^2}{\partial k_a \partial k_b}\left(k_{0(1)}
    k_{0(2)}\right)\,,
\end{eqnarray}
where we used (\ref{solk1+k2}). Therefore,
\begin{eqnarray}
  k_{0(1)} k_{0(2)} & = & \left[(2 q - 1) \mathfrak{g}^{(0 \ldots 0 ab)} -
    2 \mathfrak{g}^{(0 \ldots 0 a)} \mathfrak{g}^{(0 \ldots 0
      b)}\right] k_a k_b \nonumber\\
& =: & g^{ab} k_a k_b \, . \label{solk1k2}
\end{eqnarray}

If we substitute (\ref{solk1+k2}) and (\ref{solk1k2}) into
(\ref{Riemlightcone}), then merely a Riemannian metric shows up,
\begin{eqnarray} 
& & k_0^2 - \left(k_{0(1)} + k_{0(2)}\right) k_0 +
k_{0(1)} k_{0(2)} \nonumber\\ 
& & \qquad \qquad = k_0^2 + 2 g^{0a}k_0 k_a + g^{ab} k_a k_b \qquad\qquad \nonumber\\
& & \qquad \qquad = g^{ij} k_i k_j\,,
\end{eqnarray}
with $g^{00} = 1$. No Finslerian metric does occur. The underlying
metric $g^{ij}$ has to be of signature $\pm 2$. Otherwise it would
not lead, for prescribed $k_a$, to two real solutions $k_0$. Again,
the metric $g^{ij}$ has to possess the signature $(+1,-1,-1,-1)$.
$\blacksquare$

\section{Discussion}

As our main result, we have shown that radiative vacuum solutions of
the general Maxwell equations that do not show birefringence define
--- up to a scale transformation --- a Riemannian metric. Thus, the
requirement of vanishing birefringence automatically yields a
Riemannian structure.  No Finslerian metric can be introduced in
this way.  As a consequence, no intrinsic anisotropy in the
propagation of light can be found (intrinsic in the sense of using
merely the Maxwell equations). It is always possible to make a
coordinate transformation to a locally Minkowskian frame. This
applies also to a hypothetical higher order version of the
generalized inhomogeneous Maxwell equation like $\partial_j
(\chi^{ijkl} F_{kl}/2) + \partial_j \partial_m (\chi^{ijmkl} F_{kl})
= J^i$. Only if non--Minkowskian coordinates are related to or fixed
by other physical phenomena, then one may speak about an anisotropy
of the speed of light. Such phenomena may be related to quantum
matter described by some Dirac--like equation. Accordingly, this
anisotropy is defined {\it
  only\/} with respect to another physical phenomenon.

This situation is, of course, present in current tests searching for
an {\it anisotropy of the propagation of light}, like the modern tests
using optical cavities \cite{Muelleretal03c}.  In these tests the
isotropy of the velocity of light is tested with respect to the length
of the cavity. This length is determined by the Dirac equation but, in
part, also by the Maxwell equations. However, it turns out that for the
used materials the latter influence the length of the cavity only
marginally so that the length is mainly determined by the Dirac
equation. Therefore, {\it Michelson--Morley tests are tantamount to a
  comparison of the Maxwell with the Dirac equation}.

This result also shows that the generalized Maxwell equations alone
cannot cover the anisotropy effects of light described in the
kinematical framework of Robertson--Mansouri--Sexl
\cite{Robertson49,MansouriSexl77,MansouriSexl77a}. In the same way as
in this kinematical framework, one has to make a comparison between
the propagation of light and a length standard. This length standard
is given as such within the Robertson--Mansouri--Sexl framework. In
the present framework of dynamical test theories, this is replaced by
a comparison of the Maxwell and the Dirac equation. In this sense, one
may take the framework including a generalized Maxwell {\it and\/} a
generalized Dirac equation as the dynamical replacement for the old
Robertson--Mansouri--Sexl framework. However, one may want to go
further to the appreciably more general {\it standard model
  extension\/} (SME) of Kosteleck\'y and collaborators
\cite{Kostelecky04}, which contains more than a single generalized
Dirac equation. In any case, the birefringence of light and also of
Dirac matter waves in vacuum is truly beyond the
Robertson--Mansouri--Sexl scheme, but is included in the SME of
Kosteleck\'y.

Our main result only relies on the fourth order Fresnel equation
(\ref{Fresnel}). All propagation phenomena which lead to
characteristic equations of fourth order lead to a Riemannian metric
if one does not allow birefringence. Therefore, this also applies to
the characteristics of a generalized Dirac equation where the
$\gamma$--matrices are not assumed to fulfill a Clifford algebra. If
the Dirac characteristics do not show birefringence, then we can
conclude that the $\gamma$--matrices will fulfill a Clifford algebra.
This also follows from our general result in Sec.\ref{Finsler}.

Furthermore, our result can also be applied to the WKB approximation
of generalized particle field equation as, e.g., the generalized Dirac
equation
\cite{Laemmerzahl90,AudretschLaemmerzahl93,ColladayKostelecky98}. As a
result, one arrives at a scalar--vector--tensor theory where the
dispersion relation induces a splitting of
the mass shells according to $0 = k_0^2 - g^{ab} (p_a + \alpha_a) (p_b +
\alpha_b) + \alpha^2$. The equation of motion for the corresponding
point particle is that of a charged particle in Riemannian spacetime
with a position and time dependent mass.

\begin{acknowledgments}

We would like to thank H.~Dittus, A.~Garcia, A. Kosteleck\'y,
A.~Mac\'{\i}as, H.~M\"uller, Yu.~Obukhov, G.~Rubilar, and
S.~Tertychniy for fruitful discussions. FWH thanks H.~Dittus for the
hospitality at ZARM. CL thanks A.\ Kosteleck\'y also for the
invitation to the CPT'04 meeting where some of these ideas evolved.
Financial support from the German Space Agency DLR and from the DFG
(HE-528/20-1) is gratefully acknowledged.
\end{acknowledgments}


\end{document}